\documentclass[12pt]{report}
\usepackage{amssymb}
\usepackage{amsmath}

\usepackage{graphicx}
\begin{document}
\voffset = -2.9cm
\hoffset = -1.3cm
\def\itm{\newline \makebox[8mm]{}}
\def\ls{\makebox[8mm]{}}
\def\fra#1#2{\frac{#1}{#2}}
\def\fr#1#2{#1/#2}
\def\frl#1#2{\mbox{\large $\frac{#1}{\rule[-0mm]{0mm}{3.15mm} #2}$}}
\def\frn#1#2{\mbox{\normalsize $\frac{#1}{\rule[-0mm]{0mm}{3.15mm} #2}$}}
\def\frm#1#2{\mbox{\normalsize $\frac{#1}{\rule[-0mm]{0mm}{2.85mm} #2}$}}
\def\frn#1#2{\mbox{\normalsize $\frac{#1}{\rule[-0mm]{0mm}{3.15mm} #2}$}}
\def\hs#1{\mbox{\hspace{#1}}}
\def\b{\begin{equation}}
\def\e{\end{equation}}
\def\arccot{\mbox{arccot}}
\vspace*{6mm}
\makebox[\textwidth][c]
{\large \bf{FRW Universe Models in Conformally Flat Spacetime Coordinates}}
\vspace*{-1.0mm} \newline
\hspace*{5.0cm} {\large \bf{I: General Formalism}}
\vspace{4mm} \newline
\makebox[\textwidth][c]
{\normalsize \O yvind Gr\o n$^{*}$ and Steinar Johannesen$^*$}
\vspace{1mm} \newline
\makebox[\textwidth][c]
{\scriptsize $*$ Oslo University College, Department of Engineering,
P.O.Box 4 St.Olavs Plass, N-0130 Oslo, Norway}
%
\vspace{6mm} \newline
{\bf \small Abstract}
{\small
The $3$-space of a universe model is defined at a certain simultaneity.
Hence space depends on which time is used. We find a general formula
generating all known and also some new transformations to conformally
flat spacetime coordinates. A general formula for the recession velocity
is deduced.}
%
%
\vspace{10mm} \newline
{\bf 1. Introduction.}
\vspace{3mm} \newline
Space-time and 3-space have very different characters according
to the general theory of relativity. Spacetime is absolute,
meaning that the properties of spacetime do not depend upon the
reference frame. The $3$-space, on the other hand is relative.
Further on we will use the word {\it space} for the $3$-space defined
by a certain simultaneity.
\itm A.J.S.Hamilton and J.P.Lisle [1] have recently presented a new way
to conceptualize space in their river model of black holes. Similarly,
in a cosmological context, we here define a flow of inertial reference
particles as a continuum of free particles with a velocity field
approaching asymptotically the large scale velocity field of the
distant matter in the universe.
\itm We see that there are two different ways of thinking about space:
as a simultaneity space in $4$-dimensional spacetime or as a flow of
inertial frames. It may be useful to introduce separate names for these
conceptions. We therefore make the following definitions.
{\it Coordinate space} is a continuum of events taking place at a
constant coordinate time.
{\it Inertial flow} is a continuum of local inertial frames with
vanishing velocity at a specified position. These frames consist of
a flow of freely falling particles.
\itm The spacetime outside a black hole provides an example of both.
The coordinate space is defined by constant Schwarzschild time coordinate.
The inertial flow is a continuum of inertial frames with
vanishing velocity infinitely far from the black hole.
\itm In cosmology one uses comoving coordinates in which freely moving
reference particles have constant spatial coordinates. The coordinate
time shown on locally Einstein synchronized clocks carried by the
reference particles is called {\it cosmic time}, and the coordinate
space defined by constant cosmic time is called the {\it cosmic space}.
The cosmic inertial flow is usually called the {\it Hubble flow}.
\itm Just as one can cut many different surfaces through a 3-dimensional
body, one can define many different coordinate spaces in 4-dimensional
spacetime.
Flat spacetime, for example, can be separated in time and
space in many different ways. Two of them are:

\noindent {\Large $\cdot$} The space in Minkowski spacetime. This is the
coordinate space of a coordinate system comoving with free reference
particles that constitute a non-expanding inertial flow.

\noindent {\Large $\cdot$} The private and public space of the Milne
universe model. The private space of this universe model is just the
Euclidean space in Minkowski spacetime, and the public space is the
negatively curved coordinate space of a coordinate system comoving
with a set of free particles constituting an expanding inertial flow,
i.e. it is the cosmic space of an empty universe model.
\itm Similarly, in the non-empty Friedmann-Robertson-Walker (FRW) universe
models we shall define two types of coordinate space,

\noindent {\Large $\cdot$} cosmic space defined by constant cosmic time.

\noindent {\Large $\cdot$} conformal space defined by constant conformal time.
%
%
\vspace{5mm} \newline
{\bf 2. The line element of the FRW universe models}
\vspace{3mm} \newline
We shall here study FRW-universe models. The line element then takes
the form
\begin{equation} \label{e_64}
ds^2 = - dt^2 + a(t)^2 [\hs{0.6mm} d \chi^2 + S_k(\chi)^2 d \Omega^2]
\mbox{ .}
\end{equation}
where $d \Omega^2 = d{\theta}^2 + \sin^2 \hs{-0.5mm} \theta \hs{0.8mm}
d{\phi}^2$. Here the dimensionless quantity $\chi$ is called the standard
radial coordinate. We use a dimensionless cosmic time $t$.
The scale factor $a(t)$ is normalized so that $a(t_0) = 1$ where $t_0$
is the present age of the universe. This means that $a(t)$ represents
the distance between two objects at an arbitrary point of time relative
to their distance at the present time. The parameter $k$ characterizes
the spatial curvature with $k = 1$ for a positively curved space,
$k = 0$ for flat space, and $k = -1$ for a negatively curved space.
\itm In the case of a universe model with curved space the length
unit is equal to the present value $r_0$ of the spatial
curvature radius,
\begin{equation} \label{e_56}
r_{0} = \frac{c}{H_{0} \sqrt{|1 - \Omega_{0}|}}
\mbox{ ,}
\end{equation}
where the Hubble constant $H_0$ is the present value of the Hubble
parameter, and $\Omega_{0} \ne 1$ is the total energy density relative
to the critical density at the point of time $t_0$. This expression for
the spatial curvature follows from the Friedmann equations [2]. In a
universe model with flat space, i.e. with $\Omega_{0} = 1$, the unit
of length is equal to the Hubble length $c / H_0$ of the universe.
The unit of time is equal to the time taken by light to move a
unit of length. This implies that (except in equations \eqref{e_56}
and \eqref{e_136}) we use units so that $c = 1$.
As a consequence of curvature isotropy in space the function $S_k$
obeys [3,4]
\begin{equation} \label{e_55}
S_k \hs{-0.5mm} '^{\hs{0.5mm} 2} + k S_k^2 = 1
\mbox{ .}
\end{equation}
where ' denotes differentiation with respect to $\chi$. Hence
\begin{equation} \label{e_93}
S_k(\chi) =
\left\{ \begin{array}{lclcl}
\sin \chi  & \mbox{for} & k = 1  & , & 0 < \chi < \pi    \\
\chi       & \mbox{for} & k = 0  & , & 0 < \chi < \infty \\
\sinh \chi & \mbox{for} & k = -1 & , & 0 < \chi < \infty
\end{array} \right.
\mbox{ .}
\end{equation}
Furthermore $k = -1$ for $\Omega_0 < 1$, and $k = 1$ for $\Omega_0 > 1$.
Note that the unit of length may be written
\begin{equation} \label{e_136}
l_0 =
\left\{ \begin{array}{lcl}
\frl{c}{H_0} \hs{0.7mm} \sqrt{\frl{k}{\Omega_0 - 1}} & \mbox{for} & k \ne 0 \\
\vspace{-2.8mm} \\
\frl{c}{H_0}                                         & \mbox{for} & k = 0
\end{array} \right.
\mbox{ .}
\end{equation}
\vspace{-2mm}
\itm Each reference particle has constant radial coordinate $\chi$.
The reference particles may be thought of as clusters of galaxies.
The distance between the reference particles increases due to the
expansion of the universe. The general relativistic interpretation
is that space expands [5]. For the line element \eqref{e_64} an
elementary calculation shows that the Christoffel symbol
$\Gamma^{\chi}_{tt}$ vanishes. Hence, particles with constant $\chi$
move freely. They define an inertial flow which is the Hubble flow of
the universe model. The Hubble flow of these universe models expands
if $a(t)$ is an increasing function of $t$.
\itm The standard radial coordinate of the particle horizon is defined by
\begin{equation}\label{e_58}
\chi_H = \int_{t_i}^t \frac{dt}{a(t)}
\end{equation}
for universe models beginning at $t = t_i$, where different universe
models may have $t_i = - \infty$, $t_i = 0$ or
 $t_i > 0$.
\itm We now introduce a new time coordinate $\eta$, called the
{\it parametric time} [4], defined by
\begin{equation} \label{e_39}
\eta = \int_{t_1}^t \frac{dt}{a(t)}
\end{equation}
where $t_1$ is an arbitrary point of cosmic time corresponding to
$\eta = 0$. The line element as expressed in terms of $\eta$ takes
the form
\begin{equation} \label{e_92}
ds^2 = a(\eta)^2 \hs{0.6mm} [\hs{0.3mm} - d{\eta}^2 + d \chi^2
+ S_k(\chi)^2 d \Omega^2]
\mbox{ .}
\end{equation}
The space defined by constant $\eta$ is identical to the space defined
by constant $t$. Note that ${d\eta} / {dt} < 1$ for $t > t_0$ in expanding
universe models. Hence the parametric time is shown on clocks that slow
down with the expansion of the universe relative to the clocks showing
cosmic time, and stops in the limit $a \rightarrow \infty$.
\itm From the definitions \eqref{e_58} and \eqref{e_39} it follows that
\begin{equation} \label{e_59}
\chi_{H} = \eta
\end{equation}
for $t_i = - \infty$ and $t_i = 0$. If the universe comes into
existence at a point of time $t_i > 0$, the horizon
radius is given by
\begin{equation} \label{e_30}
\chi_{H} = \int_{t_i}^t \frac{dt}{a(t)}
= \eta - \eta (t_i)
\end{equation}
\itm The present velocity of a reference particle with
$\chi = \mbox{constant}$ with respect to the observer at $\chi = 0$
is given by Hubble's law
\begin{equation} \label{e_105}
v_H = \widehat{H}_0 a(t_0) \chi = \widehat{H}_0 \chi
\mbox{ ,}
\end{equation}
where
\begin{equation} \label{e_126}
\widehat{H}_0 = l_0 H_0
\end{equation}
is a dimensionless Hubble parameter. Note that $H_0$ has the dimension
inverse time. This means that with our dimensionless time coordinate
\begin{equation} \label{e_249}
H_0 = \frl{1}{l_0 a(t_0)} \left( \frl{da}{dt} \right)_{t=t_0}
\end{equation}
while
\begin{equation} \label{e_248}
\widehat{H}_0 = \frl{1}{a(t_0)} \left( \frl{da}{dt} \right)_{t=t_0}
\end{equation}
The velocity $v_H$ of the Hubble
flow is greater than $1$ for $\chi > 1 / \widehat{H}_0$.
\itm From equations \eqref{e_136} and \eqref{e_126} it follows that
$\widehat{H}_0 = 1$ for a flat universe. Then
\begin{equation} \label{e_250}
\widehat{H} = \frl{H}{H_0}
\end{equation}
i.e. the dimensionless Hubble parameter of a flat universe is equal to
the ratio of the ordinary Hubble parameter at an arbitrary point of time
and its present value. However, in a curved universe this normalization
of the dimensionless Hubble parameter is not compatible with the
corresponding normalization of the scale factor $a(t_0) = 1$.
\itm Note also that since all distances are scaled by $a(t)$,
the curvature radius of curved space at an arbitrary point of time is
\begin{equation} \label{e_131}
r(t) = r_0 a(t)
\end{equation}
Hence in a curved universe, the scale factor is equal to the dimensionless
curvature radius $a(t) = r(t) / r_0$.
%
%
\vspace{5mm} \newline
{\bf 3. Conformal coordinates in FRW universe models}
\vspace{3mm} \newline
The FRW universe models have vanishing Weyl curvature tensor, and
hence conformally flat spacetime. It is therefore possible to introduce
coordinates $(T,R)$ so that the line element has a conformally flat
form [6-9]. These types of coordinates were introduced in the description
of relativistic universe models by L.Infield and A.Schild [6], who
sacrificed the advantage of the standard description where freely moving
reference particles have constant coordinates, for a metric conformal to
Minkowski spacetime, where the speed of light is constant.
The coordinates $(T,R)$ are called {\it conformally flat spacetime (CFS)
coordinates} [10-12]. They have recently been applied by J.Garecki [13]
to calculate the energy of matter dominated Friedmann universes.
He has argued that in this connection there is a definite
advantage in using CFS coordinates. Furthermore, G.U.Varieschi [14] has
described universe models based on so called conformal gravity using CFS
coordinates.
\itm We shall here investigate the FRW universe models with reference to
these coordinates. Then the line element has the form
\begin{equation} \label{e_67}
ds^2 = A(T,R)^2 (-dT^2 + dR^2 + R^2 d \Omega^2)
= A(T,R)^2 ds_M^2
\mbox{ ,}
\end{equation}
where $A(T,R)$ is the CFS scale factor, and $ds_M^2$ is the line element
of the Minkowski spacetime. M.Ibison [15], K.Shankar and B.F.Whiting [16]
and M.Iihoshi et al. [9] have shown that a general coordinate
transformation that takes the line element \eqref{e_92} into the
form \eqref{e_67} is
\begin{equation} \label{e_68}
T = \frl{1}{2} \hs{0.6mm} [\hs{0.3mm} f(\eta + \chi) + g(\eta - \chi)
\hs{0.3mm}]
\mbox{\hspace{2mm} , \hspace{3mm}}
R = \frl{1}{2} \hs{0.6mm} [\hs{0.3mm} f(\eta + \chi) - g(\eta - \chi)
\hs{0.3mm}]
\mbox{ ,}
\end{equation}
where $f$ and $g$ are functions that must satisfy an identity deduced below.
The transformation \eqref{e_68} can be described as a composition of three
simple transformations. The first transforms from the coordinates $\eta$
and $\chi$ in the line element \eqref{e_92} to light cone coordinates
(null coordinates)
\begin{equation} \label{e_57}
u = \eta + \chi
\mbox{\hspace{2mm} , \hspace{3mm}}
v = \eta - \chi
\mbox{ .}
\end{equation}
This rotates the previous coordinate system by $- \pi / 4$ and scales it
by a factor $\sqrt{2}$. The scaling is performed for later convenience.
The second transforms $u$ and $v$ to the coordinates
\begin{equation} \label{e_70}
\tilde{u} = f(u)
\mbox{\hspace{2mm} , \hspace{3mm}}
\tilde{v} = g(v)
\mbox{ .}
\end{equation}
Finally, we scale and rotate with the inverse of the
transformation \eqref{e_57},
\begin{equation} \label{e_120}
T = \frl{\tilde{u} + \tilde{v}}{2}
\mbox{\hspace{2mm} , \hspace{3mm}}
R = \frl{\tilde{u} - \tilde{v}}{2}
\mbox{ .}
\end{equation}
Note that
\begin{equation} \label{e_122}
T^2 - R^2 = \tilde{u} \tilde{v}
\mbox{ .}
\end{equation}
Taking the differentials of $T$ and $R$ we get
\begin{equation} \label{e_121}
-dT^2 + dR^2 = -d \tilde{u} d \tilde{v}
= -f'(u) g'(v) \hs{0.5mm} du \hs{0.5mm} dv
= f'(u) g'(v) (-d \eta^2 + d \chi^2)
\mbox{ .}
\end{equation}
Comparing the expressions \eqref{e_92} and \eqref{e_67} for the line
element and using the previous formula, we find
\begin{equation} \label{e_90}
A(T,R)^2 = \frl{a(\eta)^2}{f'(u) g'(v)}
\end{equation}
and
\begin{equation} \label{e_81}
f'(u) g'(v) S_k(\chi)^2 = R^2
\mbox{ .}
\end{equation}
By \eqref{e_68} and \eqref{e_57} the last equation may be written as
\begin{equation} \label{e_94}
f'(u) g'(v) \hs{0.6mm} S_k \hs{-0.5mm} \left( \frl{u - v}{2} \right)^2
= \frl{1}{4} \hs{0.6mm} [ \hs{0.3mm} f(u) - g(v) \hs{0.3mm} ]^2
\mbox{ .}
\end{equation}
Substituting $v = u$ and utilizing that $S_k(0) = 0$, this equation
gives $g(u) = f(u)$. Hence equation \eqref{e_94}
reduces to
\begin{equation} \label{e_212}
f'(u) f'(v) \hs{0.6mm} S_k \hs{-0.5mm} \left( \frl{u - v}{2} \right)^2
= \frl{1}{4} \hs{0.6mm} [ \hs{0.3mm} f(u) - f(v) \hs{0.3mm} ]^2
\mbox{ ,}
\end{equation}
and equation \eqref{e_68} takes the form
\begin{equation} \label{e_211}
T = \frl{1}{2} \hs{0.6mm} [\hs{0.3mm} f(\eta + \chi) + f(\eta - \chi)
\hs{0.3mm}]
\mbox{\hspace{2mm} , \hspace{3mm}}
R = \frl{1}{2} \hs{0.6mm} [\hs{0.3mm} f(\eta + \chi) - f(\eta - \chi)
\hs{0.3mm}]
\mbox{ .}
\end{equation}
\vspace{-9.5mm} \newline
\itm Inserting $\chi = 0$ gives $T = f(\eta)$ and $R = 0$. Hence the
physical interpretation of the function $f$ is that it represents the
transformation from parametric time to conformal time at $\chi = R = 0$.
Different choices of the function $f$ generate different types of
conformal coordinates. The function $f$ is assumed to be increasing,
meaning that the conformal time proceeds in the same direction as the
parametric time at $\chi = R = 0$, i.e. in the same direction as the
cosmic time. We shall later introduce several generating functions $f$
satisfying this relationship. Inserting equation \eqref{e_81} in
\eqref{e_90} and demanding that $A(T,R) > 0$, we obtain the following
expression for the CFS scale factor
\begin{equation} \label{e_123}
A(T,R) = \frl{a(\eta(T,R)) \hs{0.5mm} S_k(\chi(T,R))}{|R|}
\mbox{ .}
\end{equation}
\vspace{-4.5mm}
\itm We shall now integrate equation \eqref{e_212}, written in the form
\begin{equation} \label{e_194}
\frl{1}{S_k \hs{-0.5mm} \left( \frl{u - v}{2} \right)^2}
= \frl{4 f'(u) f'(v)}{[ \hs{0.3mm} f(u) - f(v) \hs{0.3mm} ]^2}
\mbox{ .}
\end{equation}
For this purpose we introduce the function
\begin{equation} \label{e_195}
I_k(x) =
\left\{ \begin{array}{lcl}
\cot x  & \mbox{for} & k = 1  \\
1 / x   & \mbox{for} & k = 0  \\
\coth x & \mbox{for} & k = -1
\end{array} \right.
\end{equation}
so that $I_k' (x) = - S_k(x)^{-2}$. Inserting $v = a$ where $a$ is an
arbitrary constant in equation \eqref{e_194} and integrating leads to
\begin{equation} \label{e_196}
f(x) - f(a) = \frl{2 f'(a)}{b + I_k \left( \frl{x - a}{2} \right)}
\mbox{ ,}
\end{equation}
where $b$ is a constant of integration. This generalises a corresponding
expression deduced by Iihoshi et al. [9] who have put $a = 0$. As will be
shown below, our more general choice gives us a new solution not
obtainable with $a = 0$.
\itm Equation \eqref{e_194} has some interesting properties. It is
invariant with respect to an additive and a multiplicative constant on
the function $f$. Hence $f(a)$ and $f'(a)$ may be chosen as arbitrary
constants. This gives the solution
\begin{equation} \label{e_197}
f(x) =  c \left[ b + I_k \left( \frl{x - a}{2} \right) \right]^{-1} + d
\end{equation}
for $x \ne a$, where $c$ and $d$ are arbitrary constants. Demanding that
the function $f$ is continuously differentiable, we must define $f(a) = d$.
In appendix A we demonstrate that this function satisfies the
equation \eqref{e_212}.
The expression \eqref{e_197} contains generating functions of
many transformations to conformally flat coordinates considered
previously, as well as new ones.
\itm We shall now show that the coordinate differentials transform as
a combination of a Lorentz transformation and a scaling.
Consider a comoving particle $P_H$ in the Hubble flow, keeping $\chi$
constant. We shall relate this particle to one at rest in the
conformal coordinate system, $R = \mbox{constant}$, by a local
Lorentz transformation. The partial derivatives of the coordinates $T$
and $R$ are
\begin{equation} \label{e_96}
\frl{\partial T}{\partial \eta} = \frl{\partial R}{\partial \chi}
= \frl{1}{2} \hs{0.6mm} [ \hs{0.3mm} f'(u) + f'(v) \hs{0.3mm} ]
\end{equation}
and
\begin{equation} \label{e_97}
\frl{\partial R}{\partial \eta} = \frl{\partial T}{\partial \chi}
= \frl{1}{2} \hs{0.6mm} [ \hs{0.3mm} f'(u) - f'(v) \hs{0.3mm} ]
\mbox{ .}
\end{equation}
The {\it recession velocity} in cosmology is the coordinate velocity
of particles with $\chi = \mbox{constant}$, i.e. it is the coordinate
velocity of the Hubble flow. Obviously, the recession velocity of
the Hubble flow vanishes in the cosmic coordinate system. G.Endean has
introduced the recession velocity in the CFS coordinate system [10].
It is given by
\begin{equation} \label{e_98}
V = \tanh \theta
= \left( \frl{dR}{dT} \right)_{\chi = \mbox{\scriptsize constant}}
= \frl{f'(u) - f'(v)}{f'(u) + f'(v)}
\mbox{ .}
\end{equation}
where $\theta$ is the rapidity of the particle $P_H$ in the CFS system.
It may be noted from equation \eqref{e_98} that this velocity cannot
exceed that of light. From this equation it also follows that the
Lorentz factor is
\begin{equation} \label{e_103}
\gamma = \cosh \theta = \frl{f'(u) + f'(v)}
{2 \hs{0.5mm} [\hs{0.2mm} f'(u) f'(v) \hs{0.3mm}]^{1/2}}
\mbox{ ,}
\end{equation}
and
\begin{equation} \label{e_116}
\gamma V = \sinh \theta = \frl{f'(u) - f'(v)}
{2 \hs{0.5mm} [\hs{0.2mm} f'(u) f'(v) \hs{0.3mm}]^{1/2}}
\mbox{ .}
\end{equation}
Hence
\begin{equation} \label{e_117}
[f'(u) f'(v)]^{\hs{0.3mm} -1/2} \hs{0.5mm} dT
= \cosh \theta \hs{0.5mm} d \eta
+ \sinh \theta \hs{0.5mm} d \chi
\end{equation}
and
\begin{equation} \label{e_118}
[f'(u) f'(v)]^{\hs{0.3mm} -1/2} \hs{0.5mm} dR
= \sinh \theta \hs{0.5mm} d \eta
+ \cosh \theta \hs{0.5mm} d \chi
\mbox{ .}
\end{equation}
The differentials of the coordinates used by the two observers
are therefore related by
\begin{equation} \label{e_119}
\left[ \begin{array}{c}
dT \\
dR
\end{array} \right]
= \left[ \begin{array}{cc}
B & 0 \\
0 & B
\end{array} \right]
\left[ \begin{array}{cc}
\cosh \theta & \sinh \theta \\
\sinh \theta & \cosh \theta
\end{array} \right]
\left[ \begin{array}{c}
d \eta \\
d \chi
\end{array} \right]
\mbox{ ,}
\end{equation}
which corresponds to a composition of a Lorentz transformation
with rapidity $\theta$ and a scaling with the factor
\begin{equation} \label{e_124}
B = [f'(u) f'(v)]^{1/2} = \frl{a}{A}
\mbox{ ,}
\end{equation}
which we call the {\it relative scale factor}. It can also be expressed
in terms of $T$ and $R$ by
\begin{equation} \label{e_125}
B = \frl{|R|}{S_k(\chi(T,R))}
\mbox{ .}
\end{equation}
\vspace{-8mm} \newline
\itm Note that the scale factor $A(T,R)$ in the line element for
the CFS system depends upon the radial coordinate $R$. Hence as
described in the CFS coordinates the universe looks inhomogeneous.
This has the following explanation. Due to the relativity of
simultaneity and the relative motion of the reference particles,
$R = \mbox{constant}$, of the CFS system and those of the cosmic
system, $\chi = \mbox{constant}$, the CFS space represents a different
simultaneity space than the cosmic space. The universe is homogeneous,
but time dependent in the cosmic system. Thus the time dependence of the
cosmic system is transformed to a time and space dependence in the CFS
system.
\itm By means of the expression \eqref{e_197} for the generating function,
one may deduce the following general formula for the recession velocity
in conformally flat space (see Appendix C),
\begin{equation} \label{e_243}
V = \left( \frl{dR}{dT} \right)_{\chi = \mbox{\scriptsize constant}}
= \frac{f'(f^{-1}(T + R)) - f'(f^{-1}(T - R))}
{f'(f^{-1}(T + R)) + f'(f^{-1}(T - R))}
\end{equation}
where
\begin{equation} \label{e_244}
f'(f^{-1}(x)) = \frl{1}{2c} [ \{ b (x - d) - c \}^2 + k (x - d)^2]
\mbox{ .}
\end{equation}
Inserting the expression \eqref{e_244} into equation \eqref{e_243}, we
obtain
\begin{equation} \label{e_245}
V = \frac{2R \hs{0.3mm} [(b^2 + k)(T - d) - bc]}
{(b^2 + k) \hs{0.3mm} [(T - d)^2 + R^2] - c \hs{0.3mm} [2b(T - d) - c]}
\mbox{ .}
\end{equation}
\itm From equation \eqref{e_248} and the relation $dt = a d\eta$ it
follows that
\begin{equation} \label{e_251}
\widehat{H} = \frl{1}{a^2} \frl{da}{d \eta}
\mbox{ .}
\end{equation}
It will later be shown that the parametric time $\eta$ may be used as
a CFS time coordinate for flat universe models. This motivates the
following definition of the Hubble parameter in an arbitrary CFS
coordinate system
\begin{equation} \label{e_252}
H_R = \frl{1}{A^2} \frl{\partial A}{\partial T}
\mbox{ .}
\end{equation}
This definition means that $H_R$ represents the expansion or the
contraction of the CFS flow defined by reference particles with
$R = \mbox{constant}$. It should be noted, however, that these
particles are not freely falling. They do not constitute an
inertial flow.
\itm From equation \eqref{e_98} we find that the Doppler shift factor due
to the recession velocity is
\begin{equation} \label{e_146}
D = \sqrt{\frl{1 + V}{1 - V}} = \sqrt{\frl{f'(u)}{f'(v)}}
\mbox{ .}
\end{equation}
Using the expression \eqref{e_197} for the generating function
we arrive at the following general expression of the Doppler
factor
\begin{equation} \label{e_260}
D(T,R) = \left\{ \frl{k ( \hs{0.5mm} T + R - d \hs{0.5mm} )^2 + \hs{0.5mm}
[ \hs{0.5mm} b \hs{0.5mm} ( \hs{0.5mm} T + R - d \hs{0.5mm} )
- c \hs{0.5mm} ]^2}
{k ( \hs{0.5mm} T - R - d \hs{0.5mm} )^2 + \hs{0.5mm}
[ \hs{0.5mm} b \hs{0.5mm} ( \hs{0.5mm} T - R - d \hs{0.5mm} )
- c \hs{0.5mm} ]^2} \right\}^{1/2}
\mbox{ .}
\end{equation}
The total redshift $z$ of an object at $R$ emitting light at a time $T$
as observed at $R = 0$ at a time $T_0$ is given by [6]
\begin{equation} \label{e_147}
1 + z = D(T,R) \hs{1.0mm} \frl{A(T_0,0)}{A(T,R)}
\end{equation}
where $R = T_0 - T$, since light moves with constant velocity in the
CFS system.
%
%
\vspace{5mm} \newline
{\bf 4. Conclusion}
\vspace{3mm} \newline
The main result of the present paper is the general
expression \eqref{e_197} generating all transformations of the
form \eqref{e_68} from FRW standard coordinates to CFS coordinates.
The physical interpretation of these transformations is given
in \eqref{e_119}, showing that is it a composition of a Lorentz
transformation and a scaling. An advantage of the conformally flat form
of the line element is that the coordinate velocity of light is isotropic
and has the constant value $c$ everywhere.
\itm Furthermore, in appendix B we show how one can construct conformal
transformations between FRW spacetimes of arbitrary spatial curvature.
Specifically we may want to transform an arbitrary form of the line
element of a FRW universe to a form with a conformal factor times the
line element of a static FRW universe with scale factor equal to 1 and
a specified curvature. One may always start the construction by
introducing a parametric time. The result is a line element equal to
a conformal factor times a static FRW universe with the same spatial
curvature. The curvature can then be changed using appendix B in the
following way. First one transforms to a conformally flat form. Then
one transforms to the static form with the chosen spatial curvature
using the inverse of one of the transformations obtained by the
formula \eqref{e_197}. An important application is in the construction
of Penrose diagrams utilizing a form of the line element with a
conformal factor times the line element of Einstein's static universe
having positive spatial curvature.
\itm Further applications of the formalism developed in the present
paper will be given in later articles in this series.
%
%
%
\vspace{5mm} \newline
{\bf Appendix A. Proof that $f$ given in equation
\eqref{e_197} satisfies equation \eqref{e_212}}
\vspace{3mm} \newline
We first introduce the function
\begin{equation} \label{e_198}
C_k(x) = I_k(x) \hs{0.4mm} S_k(x) =
\left\{ \begin{array}{lcl}
\cos x  & \mbox{for} & k = 1  \\
1       & \mbox{for} & k = 0  \\
\cosh x & \mbox{for} & k = -1
\end{array} \right.
\end{equation}
so that $S_k' (x) = C_k(x)$. This function also satisfies the
relation
\begin{equation} \label{e_199}
S_k (u-v) = S_k(u) \hs{0.4mm} C_k(v) - C_k(u) \hs{0.4mm} S_k(v)
\mbox{ .}
\end{equation}
\vspace{-7mm} \newline
Furthermore
\begin{equation} \label{e_200}
f'(x) = \frl{c}{2} \left[ b \hs{0.6mm} S_k \left( \frl{x - a}{2} \right)
+ C_k \left( \frl{x - a}{2} \right) \right]^{-2}
\mbox{ .}
\end{equation}
\vspace{-2mm} \newline
Inserting the function $f$ in equation \eqref{e_212} leads to
\begin{eqnarray} \label{e_201}
\frl{[f(u) - f(v)]^2}{4 f'(u) f'(v)} &=&
\left\{ \left[ b + I_k \left( \frl{u - a}{2} \right) \right]^{-1}
- \left[ b + I_k \left( \frl{v - a}{2} \right) \right]^{-1} \right\}^2
\nonumber \\
& & \left[ b + I_k \left( \frl{u - a}{2} \right) \right]^{2}
\left[ b + I_k \left( \frl{v - a}{2} \right) \right]^{2}
S_k \left( \frl{u - a}{2} \right)^2
S_k \left( \frl{v - a}{2} \right)^2
\rule[-0.0mm]{0mm}{6.5mm}
\nonumber \\
&=& \left[ I_k \left( \frl{v - a}{2} \right)
- I_k \left( \frl{u - a}{2} \right) \right]^{2}
S_k \left( \frl{u - a}{2} \right)^2
S_k \left( \frl{v - a}{2} \right)^2
\rule[-0.0mm]{0mm}{7.5mm}
\mbox{ .}
\end{eqnarray}
\vspace{-2mm} \newline
Using the relation \eqref{e_199} we obtain
\begin{equation} \label{e_202}
\frl{[f(u) - f(v)]^2}{4 f'(u) f'(v)} =
\left[ S_k \left( \frl{u - a}{2} \right) C_k \left( \frl{v - a}{2} \right)
- C_k \left( \frl{u - a}{2} \right) S_k \left( \frl{v - a}{2} \right)
\right]^2 = S_k \left( \frl{u - v}{2} \right)^2
\end{equation}
\vspace{-2mm} \newline
which was to be shown.
\newpage
%
%
%
{\bf Appendix B. Composition of generating functions}
\vspace{3mm} \newline
In section 3 we have seen that it is possible to transform
away spatial curvature. In the same way we obtain a more general
transformation from spaces with curvature $k_1$ to spaces with
curvature $k_2$ if the generating function $f$ satisfies the relation
\begin{equation} \label{e_132}
f'(u) f'(v) \hs{0.6mm} S_{k_1} \hs{-0.5mm} \left( \frl{u - v}{2} \right)^2
= S_{\hs{0.5mm} k_2} \hs{-0.5mm} \left( \frl{f(u) - f(v)}{2} \right)^2
\mbox{ .}
\end{equation}
\vspace{-4mm}
\itm We shall now prove the following rule of composition for generating
functions. Let $f$ be a generating function between spaces with
curvatures $k_1$ and $k_2$, and $g$ a generating function between
spaces with curvatures $k_2$ and $k_3$, both satisfying equation
\eqref{e_132}. Then the composition $h$ of $g$ and $f$, $h(x) = g(f(x))$,
is a generating function between spaces with curvatures $k_1$ and $k_3$.
The rule is proved by utilizing the chain rule of differentiation.
\begin{eqnarray} \label{e_213}
& & \hspace*{-8mm}
S_{k_3} \hs{-0.5mm} \left( \frl{h(u) - h(v)}{2} \right)^2
= S_{k_3} \hs{-0.5mm} \left( \frl{g(f(u)) - g(f(v))}{2} \right)^2
= g'(f(u)) \hs{0.9mm} g'(f(v)) \hs{0.6mm} S_{k_2} \hs{-0.5mm}
\left( \frl{f(u) - f(v)}{2} \right)^2
\nonumber \\
& & \hspace*{-8mm}
= g'(f(u)) \hs{0.9mm} g'(f(v)) \hs{0.9mm} f'(u) \hs{0.6mm} f'(v)
\hs{0.9mm} S_{k_1} \hs{-0.5mm}
\left( \frl{u - v}{2} \right)^2
= h'(u) \hs{0.6mm} h'(v) \hs{0.9mm} S_{k_1} \hs{-0.5mm}
\left( \frl{u - v}{2} \right)^2
\rule[-0.0mm]{0mm}{6.5mm}
\mbox{ .}
\end{eqnarray}
\vspace{-9.5mm} \newline
\itm We also have the following rule for the inverse of a generating
function. Let $f$ be an invertible generating function between spaces
with curvatures $k_1$ and $k_2$ satisfying equation \eqref{e_132}.
Then the inverse funtion $g = f^{-1}$ is a generating function between
spaces with curvatures $k_2$ and $k_1$. For if $U = f(u)$ and $V = f(v)$,
then
\begin{eqnarray} \label{e_214}
& & \hspace*{-8mm}
S_{k_1} \hs{-0.5mm} \left( \frl{g(U) - g(V)}{2} \right)^2
= S_{k_1} \hs{-0.5mm} \left( \frl{u - v}{2} \right)^2
= \frl{1}{f'(u)} \hs{1.3mm} \frl{1}{f'(v)} \hs{1.3mm}
S_{k_2} \hs{-0.5mm} \left( \frl{f(u) - f(v)}{2} \right)^2
\nonumber \\
& & \hspace*{-8mm}
= g'(U) \hs{1.3mm} g'(V) \hs{1.3mm}
S_{k_2} \hs{-0.5mm} \left( \frl{U - V}{2} \right)^2
\rule[-0.0mm]{0mm}{6.5mm}
\mbox{ .}
\end{eqnarray}
%
%
%
\vspace{-2mm} \newline
{\bf Appendix C. A general formula for the recession velocity}
\vspace{3mm} \newline
The recession velocity is given in equation \eqref{e_98} which may be
written
\begin{equation} \label{e_235}
V = \left( \frl{dR}{dT} \right)_{\chi = \mbox{\scriptsize constant}}
= \frl{f'(f^{-1}(T + R)) - f'(f^{-1}(T - R))}
{f'(f^{-1}(T + R)) + f'(f^{-1}(T - R))}
\end{equation}
Hence we need to calculate the function $y = f'(f^{-1}(x))$ using
the expression \eqref{e_197}. This gives
\begin{equation} \label{e_236}
f^{-1}(x) = 2 \hs{0.5mm} I_k^{-1} \left( \frl{c}{x - d} - b \right) + a
\end{equation}
where
\begin{equation} \label{e_237}
I_k^{-1}(x) =
\left\{ \begin{array}{lcl}
\arccot \hs{0.5mm} x        & \mbox{for} & k = 1  \\
1 / x                       & \mbox{for} & k = 0  \\
\mbox{arccoth} \hs{0.5mm} x & \mbox{for} & k = -1
\end{array} \right.
\end{equation}
In order to combine this with formula \eqref{e_200} for $f'$, we
need the following formulae
\begin{equation} \label{e_238}
S_k(I_k^{-1}(x)) =
\left\{ \begin{array}{lcl}
\frl{1}{\sqrt{x^2 + 1}}        & \mbox{for} & k = 1  \\
1 / x \rule[-2mm]{0mm}{7.25mm} & \mbox{for} & k = 0  \\
\frl{1}{\sqrt{x^2 - 1}}        & \mbox{for} & k = -1
\end{array} \right.
\end{equation}
and
\begin{equation} \label{e_239}
C_k(I_k^{-1}(x)) = x \hs{0.5mm} S_k(I_k^{-1}(x))
\end{equation}
In the arguments of $S_k$ and $C_k$ of the formula \eqref{e_200}
we use that
\begin{equation} \label{e_240}
\frl{f^{-1}(x) - a}{2} = I_k^{-1} \left( \frl{c}{x - d} - b \right)
\mbox{ .}
\end{equation}
Combining this with equation \eqref{e_240} we get
\begin{equation} \label{e_241}
f'(f^{-1}(x)) = \frl{c}{2} \left( \frl{c}{x - d} \right)^{-2}
S_k \left( I_k^{-1} \left( \frl{c}{x - d} - b \right) \right)^{-2}
\mbox{ .}
\end{equation}
By means of equation \eqref{e_238} we finally obtain
\begin{equation} \label{e_242}
f'(f^{-1}(x)) = \frl{1}{2c} [ \{ b (x - d) - c \}^2 + k (x - d)^2]
\end{equation}
\vspace{-2mm} \newline
{\bf References}
%
\begin{enumerate}
\item A.J.S.Hamilton and J.P.Lisle, \textit{The river model
of black holes}, Am.J.Phys. \textbf{76}, 519 - 532 (2008).

\item \O .Gr\o n and S.Hervik, Einstein's General Theory of Relativity,
Springer, (2007), ch.11.

\item \O .Gr\o n, Lecture Notes on the General Theory of
Relativity, Springer, 2009, p 208.

\item C.F.Sopuerta, \textit{Stationary generalized Kerr-Schild spacetimes},
J.Math.Phys. \textbf{39}, 1024 - 1039 (1998).

\item \O .Gr\o n and \O .Elgar\o y, \textit{Is space expanding in the
Friedman universe models?}, Am.J. Phys. \textbf{75}, 151 - 157 (2007).
Further references on this topic are found here.

\item L.Infield and A.Schild, \textit{A New Approach to Kinematic
Cosmology}, Phys.Rev. \textbf{68}, 250 - 272 (1945).

\item G.E.Tauber, \textit{Expanding Universe in Conformally Flat
Coordinates}, J.Math.Phys. \textbf{8}, 118 - 123 (1967).

\item V.F.Mukhanov, \textit{Physical foundations of cosmology},
Cambridge University Press, (2005).

\item M.Iihoshi, S.V.Ketov and A.Morishita, \textit{Conformally Flat
FRW Metrics}, Prog.Theor. Phys. \textbf{118},
475 - 489 (2007)

\item G.Endean, \textit{Redshift and the Hubble Constant in Conformally
Flat Spacetime}, The Astrophysical Journal \textbf{434}, 397 - 401 (1994).

\item G.Endean, \textit{Resolution of Cosmological age and
redshift-distance difficulties}, Mon.Not. R.Astron.Soc. \textbf{277},
627 - 629 (1995).

\item G.Endean, \textit{Cosmology in Conformally Flat Spacetime}, The
Astrophysical Journal \textbf{479}, 40 - 45 (1997).

\item J.Garecki, \textit{On Energy of the Friedmann Universes in
Conformally Flat Coordinates}, arXiv:0708.2783.

\item G.U.Varieschi, \textit{A Kinematical Approach to Conformal Cosmology},
Gen.Rel.Grav. \textbf{42}, 929 - 974 (2010).

\item  M.Ibison, \textit{On the conformal forms of the
Robertson-Walker metric}, J.Math.Phys. \textbf{48},
122501-1 -- 122501-23 (2007).

\item K.Shankar and B.F.Whiting, \textit{Conformal coordinates for a
constant density star}, arXiv:0706.4324.

\end{enumerate}

\end{document}